\documentclass[a4paper,10pt]{article}
\usepackage{graphicx}
\usepackage{comment}
\usepackage{amssymb}
\usepackage{amsmath}
\usepackage{nicefrac}
\usepackage{graphicx}
\usepackage{dcolumn}
\usepackage{bm}
\usepackage{comment}
\usepackage{multirow}

\begin{document}

\huge

\begin{center}
A project based on Multi-Configuration Dirac-Fock calculations for plasma spectroscopy
\end{center}

\vspace{0.5cm}

\large

\begin{center}
Maxime Comet\footnote{maxime.comet@cea.fr}, Jean-Christophe Pain, Franck Gilleron and Robin Piron
\end{center}

\normalsize

\begin{center}
\it CEA, DAM, DIF, F-91297 Arpajon, France
\end{center}

\vspace{0.5cm}

\begin{abstract}
We present a project dedicated to hot plasma spectroscopy based on a Multi-Configuration Dirac-Fock (MCDF) code, initially developed by J. Bruneau. The code is briefly described and the use of the transition state method for plasma spectroscopy is detailed. Then an opacity code for local-thermodynamic-equilibrium plasmas using MCDF data, named OPAMCDF, is presented. Transition arrays for which the number of lines is too large to be handled in a DLA calculation can be modeled within the Partially Resolved Transition Array method or using the Unresolved Transition Arrays formalism in jj-coupling. An improvement of the original Partially Resolved Transition Array method is presented which gives a better agreement with Detailed Line Accounting computations. Comparisons with some absorption and emission experimental spectra are shown. Finally, the capability of the MCDF code to compute atomic data required for collisional-radiative modeling of plasma at non local thermodynamic equilibrium is illustrated. Additionally to photoexcitation, this code can be used to calculate photoionization, electron impact excitation and ionization cross-sections as well as autoionization rates in the Distorted-Wave or Close Coupling approximations. Comparisons with cross-sections and rates available in the literature are discussed. 
\end{abstract}

\section{Introduction}

The computation of atomic properties in a detailed level framework is central for many applications both for plasmas in local thermodynamic equilibrium (LTE) and out of LTE. For instance, accurate Detailed Line Accounting (DLA) calculations at LTE are necessary for the interpretation of some absorption experiments. Moreover, for Non-LTE plasma spectroscopy, the detailed rates of every process that are involved in the excitation and de-excitation of atomic states have to be taken into account properly to describe experimental spectra as shown for example in the last Non-LTE code comparison workshop \cite{Piron17,Hansen13}. Atomic-physics codes dedicated to this application, often use an Hartree-Fock-Slater central potential (or its relativistic version like ATOMIC \cite{Sampson09} and FAC \cite{Gu08}) or a parametric potential (such as in HULLAC \cite{Gilles15}). This treatment is generally satisfactory to study highly ionized species. However, some investigations have shown that the use of a Dirac-Fock versus a Dirac-Fock-Slater potential has non negligible effects on $\Delta n=0$ collisional-excitation cross-sections \cite{Zhang93}. The increase of computer performances makes possible the use of a more time-consuming formalism such as Multi Configuration Dirac-Fock (MCDF) to study the radiative properties of plasmas. Only few codes use such formalism to calculate opacity and emissivity spectra of plasmas \cite{Blancard12}.

In this paper we present a project dedicated to spectroscopy of LTE and Non-LTE plasmas using the MCDF code initially developed by J. Bruneau \cite{Bruneau83}. We start with a general presentation of the MCDF formalism and the use of the transition state method to improve results on the line energies is discussed. The OPAMCDF code which is able to compute emissivity or opacity spectra for LTE plasmas is detailed. The capability of the MCDF code to compute cross-sections and rates for Non-LTE plasma spectroscopy is presented in the last part. Some comparisons with cross-sections and rates available in the literature are finally discussed.

\section{Presentation of the MCDF code}

The MCDF code used in this work, written by J. Bruneau \cite{Bruneau83}, is based on methods published by I. P. Grant \cite{Grant70,Grant80}. The MCDF formalism being widely described in the literature \cite{Grant07}, we will only briefly recall its main features. The wavefunction of a state, $|\Phi(\Pi J M)>$, called Atomic State Function (ASF), is characterized by the total angular momentum $J$, its projection $M$ and the parity $\Pi$ expressed as a linear combination of Configuration State Function (CSF) $|\phi(\nu \Pi J M)>$:

\begin{equation}
|\Phi(\Pi_t J_t M_t)> = \sum_{i=1}^{N_{\mathrm{CSF}}} c_{it} |\phi(\nu_i \Pi_i J_i M_i)>
\end{equation} 

\noindent where $N_{\mathrm{CSF}}$ is the total number of CSF, $c_{it}$ are the mixing coefficients related to the state $t$ and $\nu$ denotes the ensemble of quantum numbers required to describe unambiguously a configuration. The energy of a state $E(\Pi J M)$ is therefore expressed as:

\begin{equation} 
E (\Pi_t J_t M_t) = \sum_{i=1}^{N_{\mathrm{CSF}}} \sum_{j=1}^{N_{\mathrm{CSF}}} c^*_{it} H_{ij} c_{jt}
\end{equation}

\noindent where $H_{ij}$ are the matrix elements of the Hamiltonian in the CSF basis. The mixing coefficients are determined by diagonalizing the Hamiltonian. To ensure orthogonality and normalization of the Radial Wave Functions (RWF), Lagrange's method is used to build the functional $F$:

\begin{equation}
F =  \sum_{r=1}^{N_{\mathrm{CSF}}} \sum_{s=1}^{N_{\mathrm{CSF}}} d_{rs} H_{rs}  + \sum_{A}^{N_{\mathrm{orb}}}\sum_{B}^{N_{\mathrm{orb}}} (1-\delta_{AB}) \lambda_{AB} \langle A|B\rangle
\end{equation} 

\noindent where $A$ and $B$ are the RWF and $\langle A|B\rangle$ the orthogonality integral, $\lambda_{AB}$ the Lagrange multipliers and $d_{rs}$ the generalized weights. The self-consistent-field (SCF) equations are then obtained by requiring that this functional is stationary with respect to small RWF variations. The MCDF method takes exactly into account the exchange contribution on wavefunctions. The generalized weights are linked to the average configuration (also named generalized configuration) for which the wavefunctions are determined. The generalized occupation number $\bar{q}_a$ of an orbital $a$ is then:

\begin{equation}
\bar{q}_a = \sum_k^{N_{\mathrm{CSF}}} d_{kk} q_a(k)
\end{equation}

\noindent where $q_a(k)$ is the occupation number of the orbital $a$ of the CSF $k$. The various methods of determining the generalized weights leads to different variants of MCDF calculations, which can be separated in three types \cite{Grant76}: 

\begin{enumerate}
\item The Optimized Level (OL): the weights are chosen in order to solve the SCF equations for one ASF (labeled $t$). The weight is then given by $d_{rs}=c_{rt} c_{st}$.
\item The Extended Optimized Level (EOL): the weights are chosen in order to solve the SCF equations for $1<N_{\mathrm{ASF}}<N_{\mathrm{max}}$ where $N_{\mathrm{ASF}}$ is the number of ASF being solved and $N_{\mathrm{max}}$ the total number of ASF. The weights are expressed as: 

\begin{equation}
d_{rs} = \frac{\sum_{i=1}^{N_{\mathrm{ASF}}}(2J_i + 1)c_{ri} c_{si}}{\sum_{i=1}^{N_{\mathrm{ASF}}}(2J_i + 1)}
\end{equation}

\noindent where $J_i$ is the total angular momentum of ASF $i$. In OL and EOL calculations, the dependence of the RWF on the mixing coefficients through the generalized weights implies that the SCF equations and the diagonalization of the Hamiltonian can not be separated.

\item The Average Level (AL): the weights are chosen in order to solve the SCF equations for all the ASF and expressed as: 

\begin{equation}
d_{rs} = \frac{\delta_{rs}(2J_r + 1)}{\sum_{i=1}^{N_{\mathrm{CSF}}}(2J_i + 1)}
\end{equation}

In this case, the weights do not depend on the mixing coefficients so the SCF equations and the diagonalization can be performed separately.

\end{enumerate}

For the study of highly ionized species, the AL method is the one that is generally used. Indeed, it gives a good estimation of all the level and line energies in only one calculation. Moreover, for the calculations of matrix elements, wavefunctions of the initial and final states are orthogonal. 

\subsection{The ``transition state'' method}

Lets assume that the degeneracy of the initial configuration of a one-electron transition is much higher than the degeneracy of the final one. In AL calculation, the few states of the final configuration will not play any role in the minimization procedure because their degeneracy is small. The wavefunctions will then describe preferentially the initial configuration. Relaxation of the wavefunctions is thus not taken into account properly and the transition energies may not be sufficiently accurate.

To improve the accuracy of AL calculations, J. Bruneau used the transition state (TS) method, originally developed by Slater, in order to equilibrate the weight between the initial and final configurations \cite{Bruneau83}. If the initial configuration is $\alpha$ and the final one $\beta$ then the weights are calculated by:

\begin{equation}
d^\alpha_{rs} = \frac{1}{2}\frac{\delta_{rs}(2J^\alpha_r + 1)}{\sum_{i=1}^{N_{\mathrm{CSF}} (\alpha)}(2J^{\alpha}_i + 1)}
\end{equation}

\begin{equation}
d^\beta_{rs} = \frac{1}{2}\frac{\delta_{rs}(2J^\beta_r + 1)}{\sum_{i=1}^{N_{\mathrm{CSF}}(\beta)}(2J^{\beta}_i + 1)}
\end{equation}

\noindent where the factor $1/2$ ensures that:

\begin{equation}
\sum_{i=1}^{N_{\mathrm{CSF}} (\alpha)} d^\alpha_{ii} + \sum_{j=1}^{\mathrm{CSF} (\beta)} d^\beta_{jj} = 1
\end{equation}

\begin{figure}[h]\begin{center}
\includegraphics[scale=0.238]{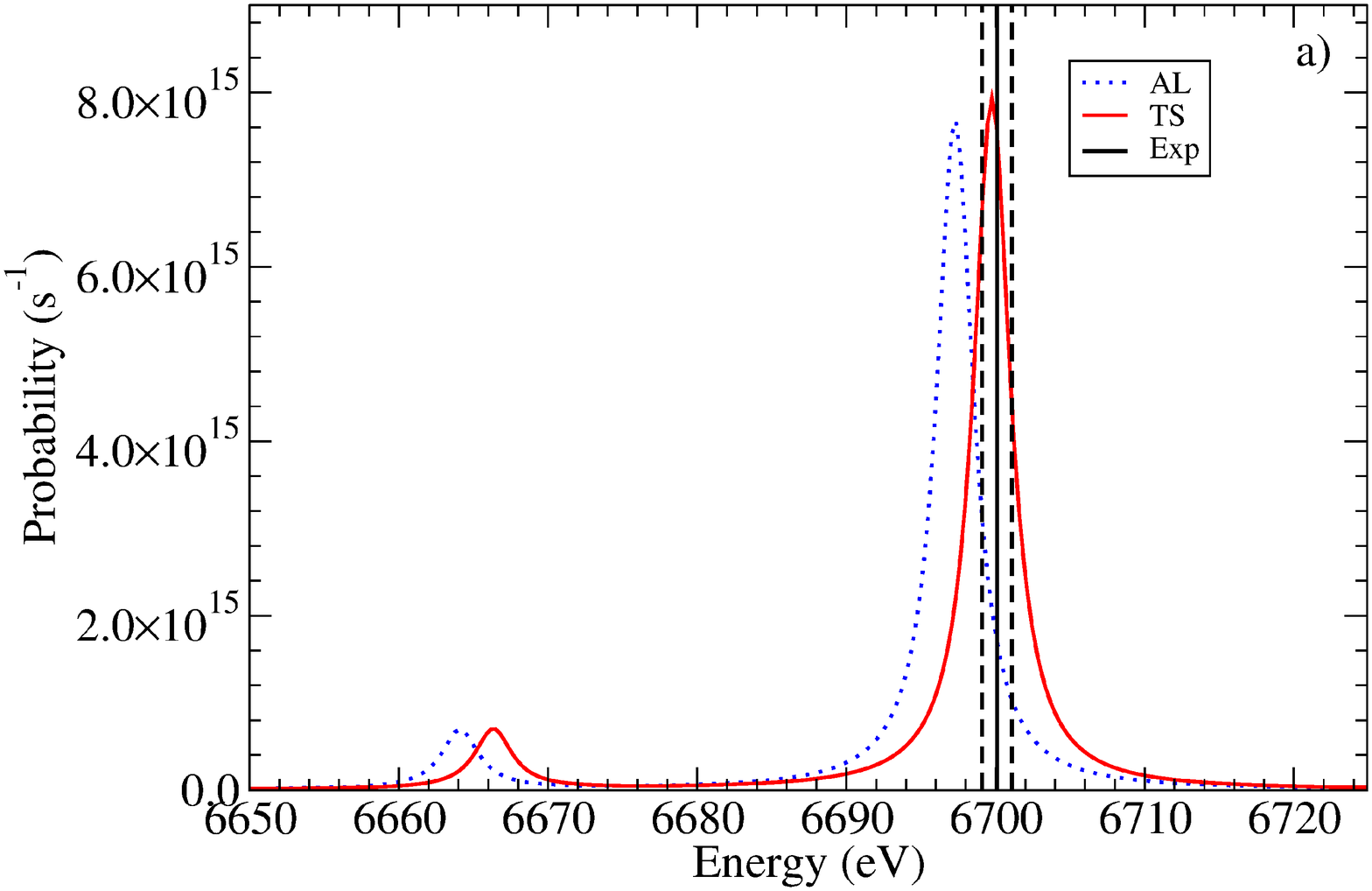}
\includegraphics[scale=0.25]{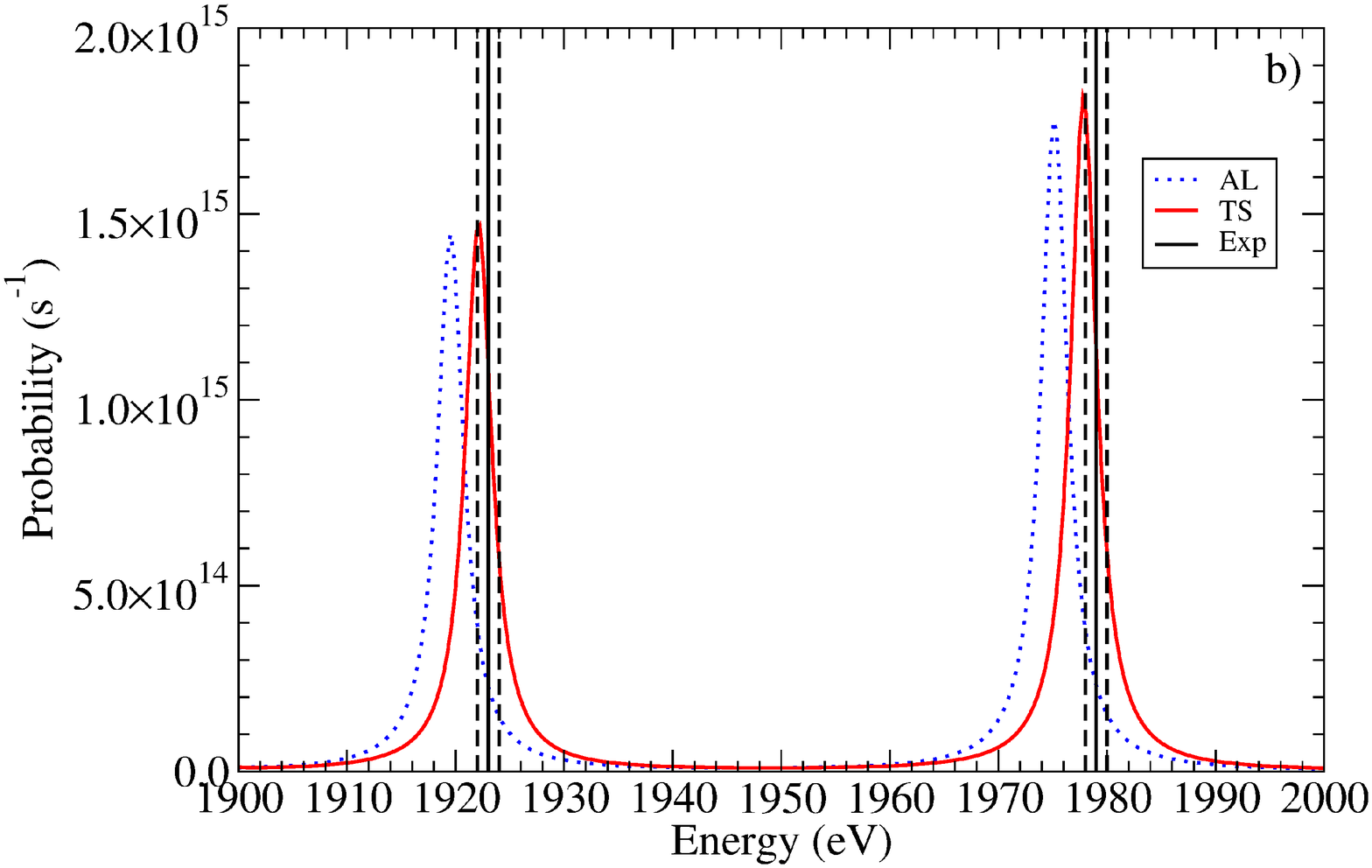}
\caption{(Color online) Comparison of the line energies of Fe$^{24+}$ 2p-1s transition (a) and Rb$^{27+}$ 3d-2p transition (b) using (red curve) or not (dotted blue curve) the transition state method. For the 2p-1s of Fe and 3d-2p of Rb, the calculated line energies are compared with the experimental values \cite{Kramida15} and \cite{Denis-Petit14} respectively (vertical lines) with a one eV uncertainty (vertical dashed lines).}
\label{Fe_Rb_TS}
\end{center}\end{figure}

To illustrate the effect of the weighting of states, Figure \ref{Fe_Rb_TS} a) and b) present the 2p-1s transition of Fe$^{24+}$ and 3d-2p of Rb$^{27+}$ where each transition has been calculated using either the AL or the TS method. Results are compared to the experimental values \cite{Kramida15,Denis-Petit14}.

When the AL method is used, lines are shifted of about 2.5 to 5 eV with respect to the experimental ones. By using the TS method, in both examples, the agreement with the experimental values is improved. This effect can also be seen on the generalized configuration (see Table \ref{table_GO}). The generalized occupation of the final orbital of the transition is higher with the TS method than with the AL.

\begin{table}
\begin{center}
\begin{tabular}{|c|c|c|}
\hline
Element & Method & Generalized configuration \\
\hline
\multirow{2}{*}{Fe$^{24+}$ 2p-1s} & AL & 1s$^{1.08}$ $2p_-^{0.31}$ 2$p_+^{0.61}$ \\

 & TS & 1s$^{1.5}$ 2$p_-^{0.17}$ 2$p_+^{0.33}$ \\
\hline 
\multirow{2}{*}{Rb$^{27+}$ 3d-2p} & AL &2p$_-^{1.67}$ 2$p_+^{3.34}$ 3$d_-^{0.39}$ 3$d_+^{0.60}$  \\
 & TS & 2p$_-^{1.83}$ 2$p_+^{3.67}$ 3$d_-^{0.2}$ 3$d_+^{0.3}$ \\

\hline
\end{tabular}
\caption{Generalized configuration determined using the AL or the TS method for the Fe$^{24+}$ 2p-1s transition and Rb$^{27+}$ 3d-2p transition.}
\label{table_GO}
\end{center}
\end{table}  

\section{The OPAMCDF code for  local-thermodynamic-equilibrium spectroscopy}

This code aims to calculate opacity and emissivity spectra in local thermodynamic equilibrium using atomic-physics data calculated using MCDF (level and line energies, oscillator strengths). The code includes configuration interaction effects within each non-relativistic configuration (CI in NRC). 

\subsection{Configurations selection} 

The configuration selection is mainly based on \cite{Comet15}. To generate the list of configurations that have to be calculated, a Relativistic Average Atom Model (RAAM) is used in order to determine average quantities of the plasma such as the average ionization and the average occupation of each subshell. The population variance of each subshell is estimated using a binomial function, assuming uncorrelated electrons:

\begin{equation}
\sigma^2_k = g_k \bar{P}_k(1-\bar{P}_k)
\end{equation}

\noindent where $g_k$ is the degeneracy of the subshell $k$ and $\bar{P}_k$ is the average occupation in the $k$ subshell provided by the RAAM calculation. Only configurations with populations ranging from Max(0, $N_k$ -  $m\sigma_k$) to 
Min($g_k$, $N_k$ +  $m\sigma_k$) are considered. The value of $m$ is generally set to 3.

Once all configurations have been generated, they are sorted according to their decreasing LTE probability $\bar{P}(C)$, calculated using the grand partition function:

\begin{equation}
\bar{P}(C) = \frac{g_C \exp \left(-\beta(E_C- \mu Q)\right)}{\sum_{i=1}^{N_{Cmax}}g_i \exp\left(-\beta(E_i- \mu Q_i)\right)}
\end{equation} 

\noindent where $g_C$ is the degeneracy of the configuration $C$, $E_C$ the configuration energy, $\mu$ the chemical potential of the plasma calculated by the RAAM, $\beta=1/(k_BT)$ and $N_{Cmax}$ the total number of configurations generated. Configuration energy E$_C$ is estimated using the average wavefunctions of the RAAM calculations:

\begin{equation}\label{E_conf}
E_C = \sum^{N_{\mathrm{orb}}}_{i=1} N_i \bar{I}_i + \frac{1}{2} \sum^{N_{\mathrm{orb}}}_{i=1} \sum^{N_{\mathrm{orb}}}_{j=1} N_i ( N_j - \delta_{ij}) \bar{V}_{ij} + \sum^{N_{\mathrm{orb}}}_{i=1} N_i \bar{V}_i
\end{equation}

\noindent where $\bar{I}_k$ is the average electron nucleus interaction plus the average kinetic energy of an electron in orbital $k$, $\bar{V}_{ij}$ is the average bound-bound interaction between electrons in the $i$ and $j$ orbitals, $\bar{V}_{i}$ is the average bound-free interaction and $N_{\mathrm{orb}}$ is the total number of subshells of the configuration. 

\subsection{Opacity components}

For the calculation of the bound-bound opacity, subshells are divided into two parts. Generally the first part contains subshells with principal quantum number ranging from 1 to 6 and the second part concerns the subshells with principal quantum number $n\geq7$. Transition arrays between configurations that have no electron in the second part are treated either in:

\begin{enumerate}
\item Detailed Level Accounting if $L \leq L_{\text{max}}$
\item Partially Resolved Transition Array (PRTA \cite{Iglesias12,Iglesias12a}) if $L > L_{\text{max}}$
\end{enumerate}

\vspace{4mm}

\noindent where $L$ is the number of lines in the transition array and $L_{\text{max}}$ is the maximum number of lines generally set to 2x10$^6$. This value can be increased to 10$^7$.

However, if there is at least one electron in the second part (highly excited configurations), transition arrays are calculated in the Unresolved Transition Arrays (UTA) formalism \cite{Bauche-Arnoult79, Bauche-Arnoult82} in jj coupling called Spin-Orbit Split Array \cite{Bauche-Arnoult85} (SOSA), relying on an efficient direct computation of the two electron relativistic energy variance and shift \cite{Krief15}. 

For each transition array between two configurations, an MCDF calculation is performed using the transition state method. As mentioned before, it ensures a good accuracy in the line energies and in the orthogonality of the initial and final wavefunctions. All MCDF calculations include the Breit interaction, QED corrections (self-energy, vacuum polarization) and nucleus effects (finite nuclear mass and recoil). Moreover, wavefunctions take into account the finite size of the nucleus using a Fermi-Dirac distribution.

In the present version of the code, the bound-free component is calculated in the Detailed Configuration Accounting approximation.

\subsection{Improvement of the Partially Resolved Transition Array method}

\subsubsection{General presentation}

Let us consider a transition array between two configurations of the form:

\begin{equation}
\eta^{N_1}_1 \lambda^{N_i}_i \lambda^{N_f}_f ~ ... ~ \eta^{N_M}_M \rightarrow \eta^{N_1}_1 \lambda^{N_i-1}_i \lambda^{N_f+1}_f~...~\eta^{N_M}_M
\end{equation}

\noindent where $\lambda_i$ and $\lambda_f$ are the initial and final active subshells and $\eta_k$ the passive subshells of the transition array. The total variance of the transition array is expressed as \cite{Iglesias12a}:

\begin{equation}
\sigma^2_{tot} = \sigma^2_{EL} + \sigma^2_{SO}(\lambda_i, \lambda_f)
\end{equation}

\noindent with $\sigma^2_{EL}$ and $\sigma^2_{SO}$ the electrostatic and spin-orbit variances respectively. The latter concerns only the active subshells $\lambda_i$ and $\lambda_f$ whereas the former can be expressed as:

\begin{eqnarray}
&\sigma^2_{EL}& = \Omega^2(\lambda_i^{N_i} \lambda_f^{N_f} \rightarrow \lambda^{N_i - 1}_i \lambda^{N_f+1}_f) + \nonumber \\
& & \sum^{M}_{j=1} \frac{N_j(g_j - N_j)}{g_j - 1} \Omega^2 (\eta_j \lambda_i \rightarrow \eta_j \lambda_f)
\end{eqnarray}

\noindent where we have distinguished the contributions of active and passive subshells to the electrostatic variance. The idea of the PRTA method is to split the calculation of a transition array between a DLA calculation and a statistical part. The DLA calculation is carried out using a configuration where some passive subshells have been removed (this configuration is called the main group or reduced configuration). All the removed subshells are called the secondary group. A width is then added to each line of the DLA calculation corresponding to the contribution of the secondary group to the variance. This separation of a configuration between two groups is exact only if the Slater's integrals of the initial and final configurations are equal. 

\vspace{4mm}

In OPAMCDF, a PRTA calculation is performed in several steps:

\begin{enumerate}
\item Determination of the passive subshells that can be removed from the "real" configuration to form the reduced configuration (see section \ref{sss:passive_subshell_prta})
\item Calculation of the wavefunctions of the "real" configuration using the configuration-average mode with the same generalized configuration that would be obtained from an TS computation.
\item The DLA computation of the reduced configuration is performed using the wavefunctions of the "real" configuration previously calculated.
\item An energy shift is then applied to the detailed transition array, coming from the passive subshells electrostatic interaction. This shift $\delta E $ is calculated by \cite{Pain15}:

\begin{equation}
\delta E_{i\rightarrow f} = \sum^{M}_{j=1} N_j (V_{jf}-V_{ji})
\end{equation}

\noindent where $V_{ij}$ represents the electrostatic interaction between the $nl$ subshells $i$ and $j$. The non-relativistic electrostatic interaction is calculated by taking the mean of the relativistic electrostatic interactions. This shift has also to be applied on the oscillator strength calculations because of their transition energy dependence.
\item The electrostatic variance due to the passive subshells is added to each line of the DLA calculation in order to keep constant the total oscillator strength of the transition array.
\end{enumerate}  

The choice of the subshells that can be removed is difficult. The natural choice consists in removing subshells that have the smallest contribution to the total variance of the transition array \cite{Iglesias12}.

\subsubsection{Passive subshell selection in OPAMCDF}
\label{sss:passive_subshell_prta}

In OPAMCDF, subshells are sorted according to their contribution to the total variance of the transition array. The criterion is based on the number of lines: subshells with the smallest contribution to the variance of the transition array are removed until the number of lines of the reduced DLA calculation is lower than $L_{\text{max}}$. In Ref. \cite{Iglesias12}, the passive subshells are removed from the "real" configuration whatever their population. In fact, this can be improved by noticing that the electrons, in a subshell, can be removed "one by one". Indeed, the contribution of a subshell $j$ to the total variance can be written as:

\begin{eqnarray}
\frac{N_j(g_j - N_j)}{g_j - 1} \Omega^2 (\eta_j \lambda_i \rightarrow \eta_j \lambda_f) &=& \left( \frac{N_a(g_j - N_a)}{g_j - 1} + \frac{N_b(g_j - N_b)}{g_j - 1} - \frac{2N_aN_b}{g_j - 1} \right)
\Omega^2 (\eta_j \lambda_i \rightarrow \eta_j \lambda_f)
\end{eqnarray}

\noindent where $N_j$ is the total number of electron in subshell $j$, $N_a$ is the number of electrons in subshell $j$ that are kept in the main group whereas $N_b$ are those removed from the subshell ($N_j=N_a+N_b$). The variance of the passive electrons removed from the subshell is then:

\begin{eqnarray}
\Omega^2 (\eta_j^{N_b} \lambda_i \rightarrow \eta_j^{N_b} \lambda_f) &=& \left( \frac{N_b(g_j - N_b)}{g_j - 1} - \frac{2N_aN_b}{g_j - 1} \right) \Omega^2 (\eta_j \lambda_i \rightarrow \eta_j \lambda_f)
\end{eqnarray}

\begin{figure}\begin{center}
\includegraphics[scale=0.245]{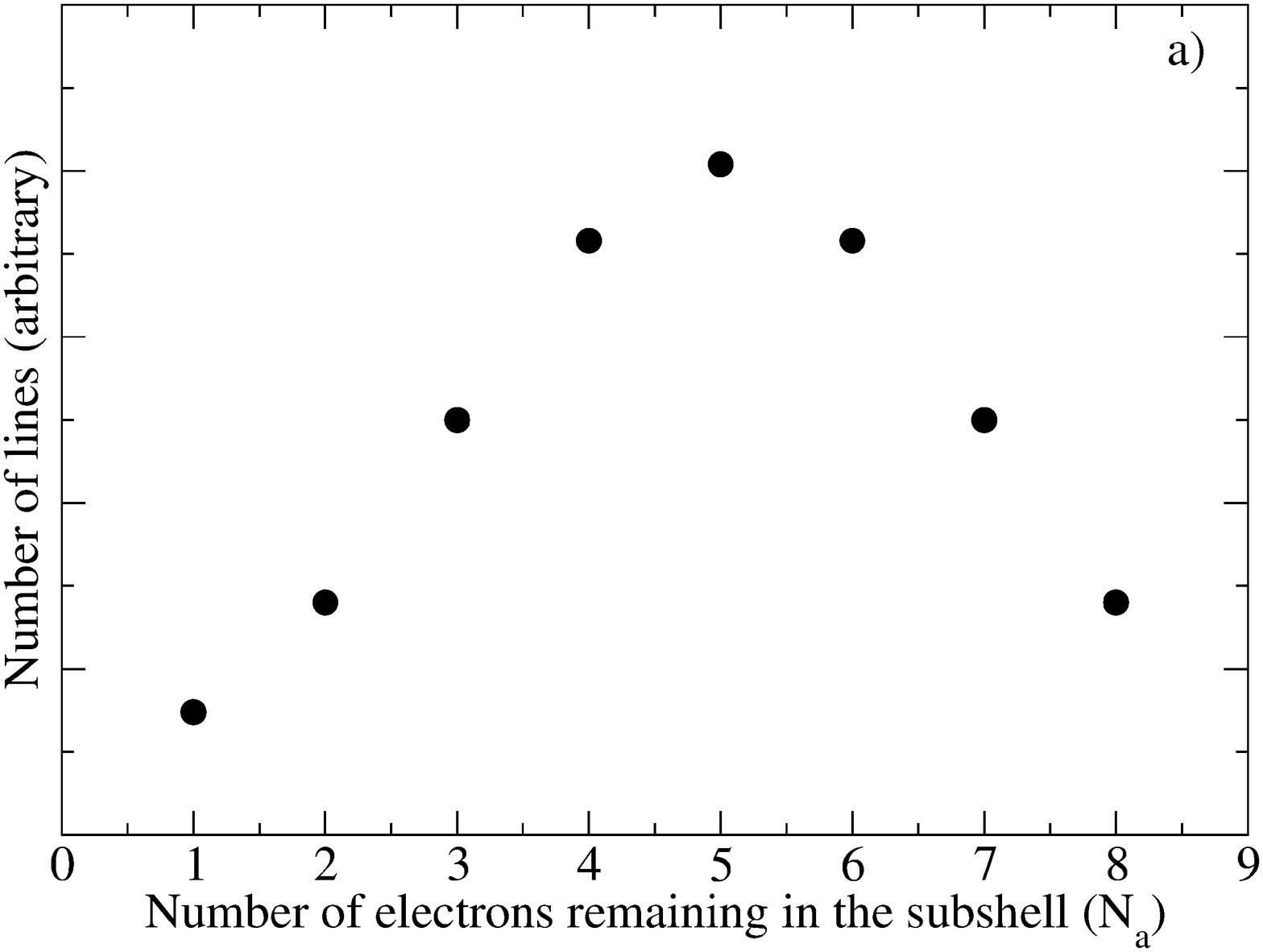}
\includegraphics[scale=0.25]{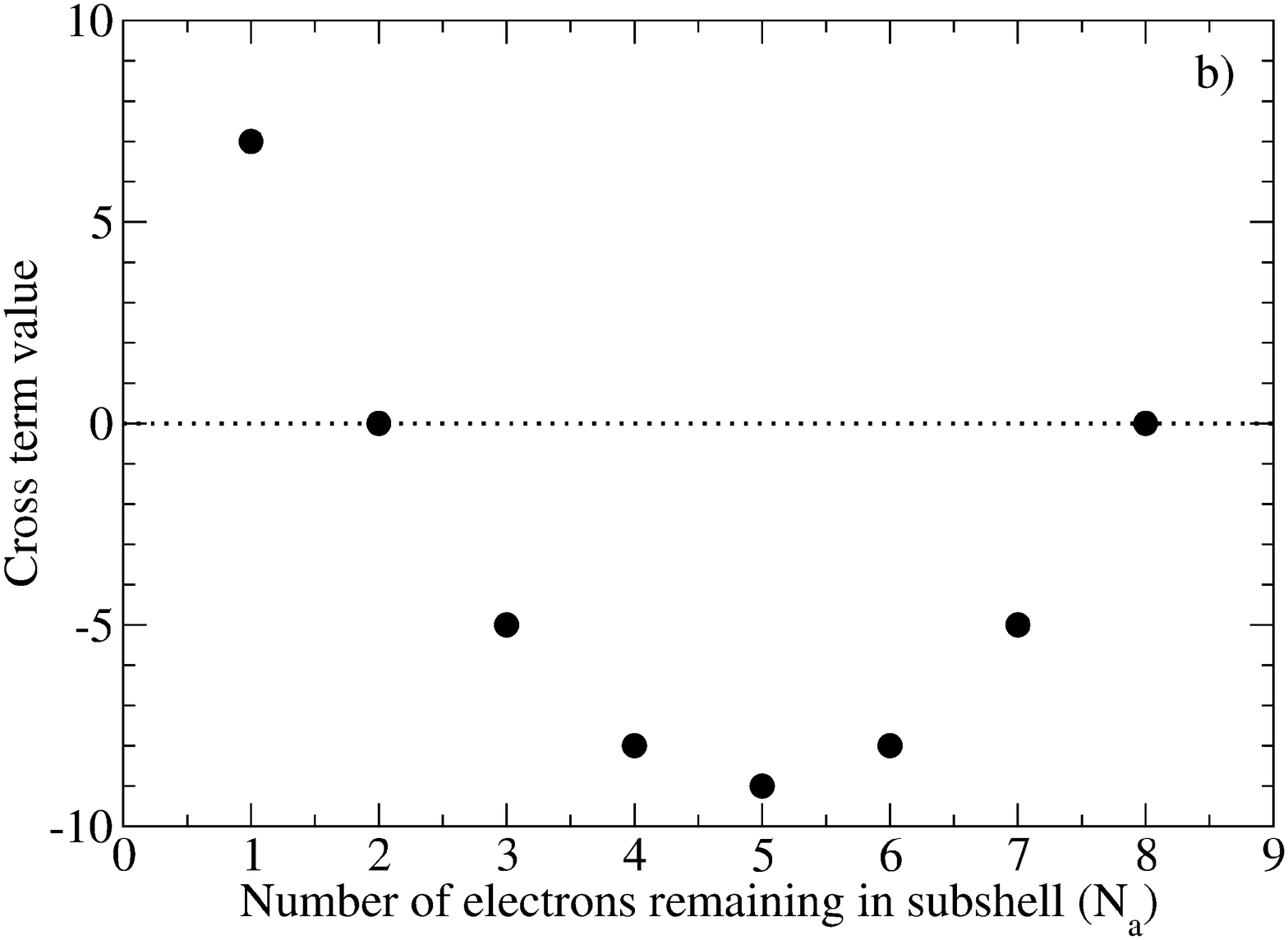}
\caption{Number of lines behavior as a function of the number of electrons remaining in a 3d subshell starting from $N_a$=8 ($N_b=0)$ (a). Cross-term value variation as a function of the number of electrons $N_a$ remaining in this subshell (b).}
\label{cross-term-example}
\end{center}\end{figure}

We have to pay attention to the fact that this variance could be negative. The variance is positive or null only if $N_a \leq g_j - N_j$. In fact, the variance due to the passive electrons in a subshell is negative only if the number of lines increase by removing this electron. To illustrate this point, lets assume an initial configuration with a 3d$^8$ passive subshell. The statistical number of lines in a transition array between configurations A and B can be estimated using a Gaussian distribution \cite{Bauche15}:

\begin{equation}
L(A-B)=\frac{3}{\sqrt{8\pi}} g_A g_B (\nu_A + \nu_B)^{-3/2}\left( 1-\frac{1}{\nu_A + \nu_B} \right) 
\end{equation} 

\noindent where $g_A$ ($g_B$) is the degeneracy of the configuration $A$ ($B$) and $\nu_A$ ($\nu_B$) are the variances of the distribution of the $\alpha JM$ states of configurations $A$ and $B$ (eq. (3.11) of \cite{Bauche15}). This expression can be simplified by assuming that ${\nu_A + \nu_B} \gg 1$ and keeping only the dependence with respect to the number of electrons $N_a$ in the passive subshell, we have:

\begin{equation}
L(A-B) \propto 
\begin{pmatrix}
g_j \\
N_a
\end{pmatrix} [N_a(g_j-N_a)]^{-3/2}
\label{eq_nb_raies}
\end{equation}
 
\noindent where $g_j$ is the degeneracy of the passive subshell $j$. In Figure \ref{cross-term-example} a) we show the number of lines calculated with eq. (\ref{eq_nb_raies}) as a function of the number of electrons remaining in the 3d subshell starting from $N_a$=8 ($N_b=0)$. This variation is compared in figure \ref{cross-term-example} b) to the cross-term value: 

\begin{eqnarray}
N_b(g_j - N_b) - 2N_aN_b = N_j(g_j - N_j) - N_a(g_j - N_a)
\label{CT}
\end{eqnarray}

\noindent where $N_j=N_a+N_b$. As displayed in Figure \ref{cross-term-example} a), starting from $N_a=8$ and removing one to six electron(s) in it, the number of lines is greater than with the original populations. In the interval $2<N_a<8$, the cross-term value is negative and null for $N_a=2$ because of the particle-hole symmetry. Then, it starts to be positive as soon as the number of lines with $N_a <8$ electrons is smaller than with $N_a=8$.

We will now illustrate this improvement with three examples, calculated for Ge at 75 eV and 0.1 g.cm$^{-3}$. All results are compared to SOSA and UTA calculations.

\subsubsection{Example 1}

\begin{figure}[h]\begin{center}
\includegraphics[scale=0.35]{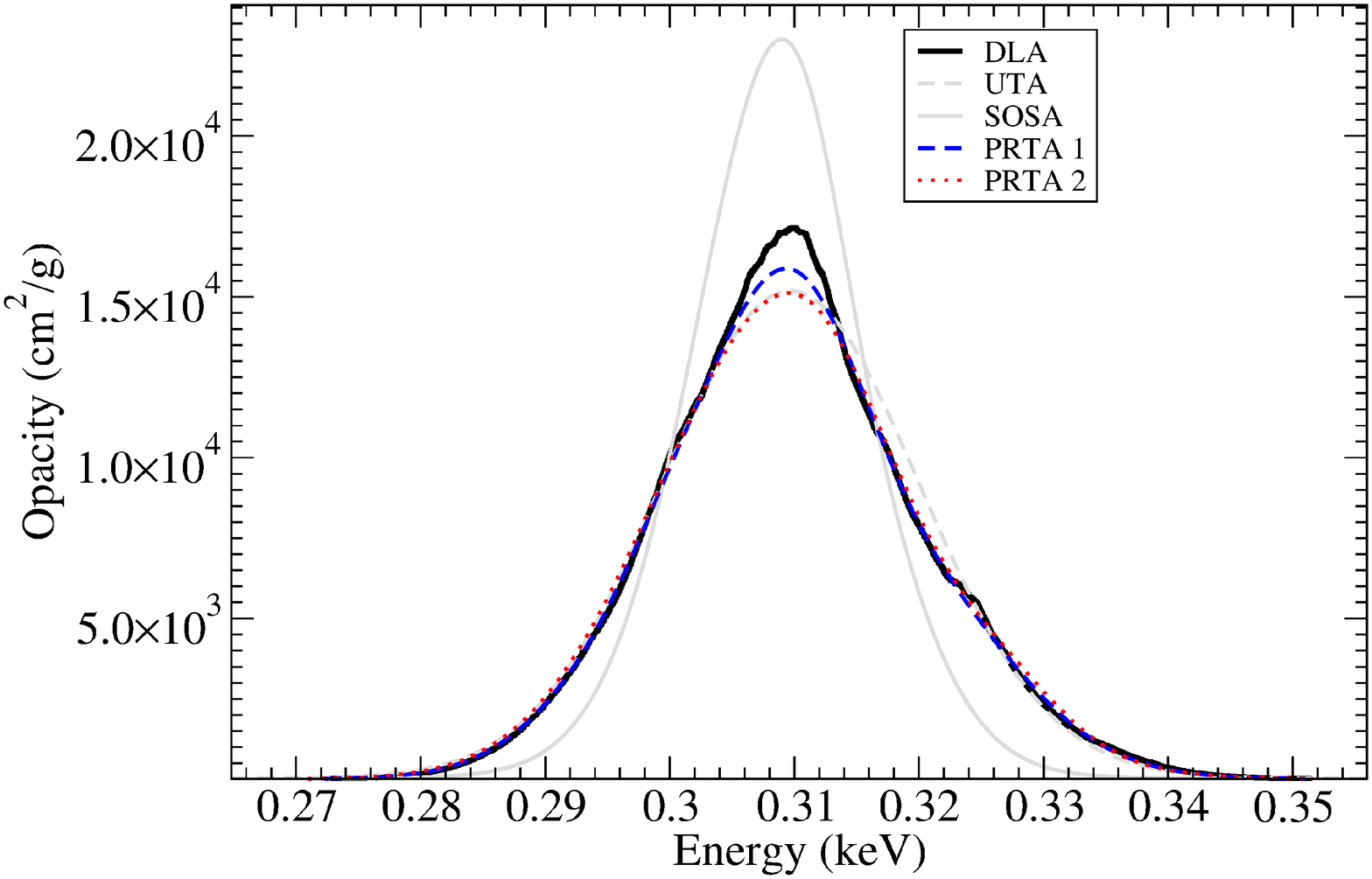}
\caption{(Color online) Comparison between the DLA and the PRTA calculations of the 3p-4d transition with an initial configuration of [Mg]3p$^3$ 3d$^2$ 4p$^2$ (example 1). The UTA and SOSA statistical modelings are also displayed.}
\label{PRTA_ex1}
\end{center}\end{figure}

The initial configuration is [Mg]3p$^3$ 3d$^2$ 4p$^2$ and the transition is 3p-4d. The DLA calculation contains approximately 1.1x10$^7$ lines. The first passive subshell with the smallest contribution to the variance is 4p. In PRTA 1 and 2, one and two electron(s) are removed from the 4p subshell respectively. The Table \ref{table_ex1} shows the main group and the number of lines of the PRTA calculations. In this example, the two PRTA calculations are very close and in good agreement with the DLA computation as we can see in Figure \ref{PRTA_ex1}. This can be easily explained by the fact that the contribution to the variance of a 4p electron is very small so removing one or two electrons does not affect strongly the transition array.

\begin{table}
\begin{center}
\begin{tabular}{|c|c|c|}\hline
PRTA & Main Group & N$_{\text{lines}}$ \\
\hline
1    & [Mg]3p$^3$ 3d$^2$ 4p$^1$ & 1.9x10$^6$ \\
2    & [Mg]3p$^3$ 3d$^2$ 4p$^0$ & 6x10$^4$ \\
\hline
\end{tabular}
\caption{Main group and number of lines of the PRTA calculations of example 1.}
\label{table_ex1}
\end{center}
\end{table}  

\subsubsection{Example 2}

\begin{figure}[h]\begin{center}
\includegraphics[scale=0.35]{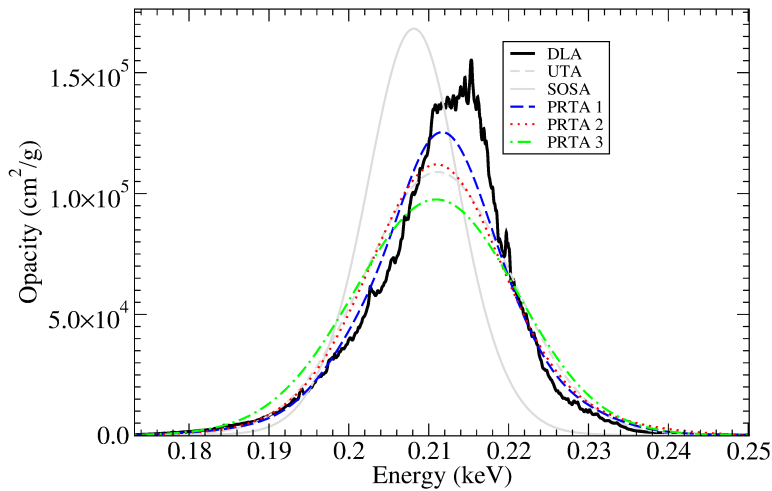}
\caption{(Color online) Comparison between the DLA and the PRTA calculations of the 3d-4f transition with an initial configuration of [Mg]3p$^3$ 3d$^5$ (example 2). The UTA and SOSA statistical modelings are also displayed.}
\label{PRTA_ex2}
\end{center}\end{figure}

The initial configuration is [Mg]3p$^3$ 3d$^5$ and the transition is 3d-4f. The DLA calculation contains 1.7x10$^6$ lines. In this example, only the 3p passive subshell contributes to the variance. In PRTA 1, 2 and 3, one, two and three electron(s) are removed from the 3p subshell respectively. Table \ref{table_ex2} shows the main group and the number of lines of the PRTA calculations. As seen in Figure \ref{PRTA_ex2}, the differences between the PRTA and DLA calculations are significant. We see that when removing only one electron in the 3p subshell, the number of lines drops to 1.7x10$^5$, leaving unchanged the asymmetrical shape of the transition array. Removing more electrons deteriorates this agreement.

\begin{table}
\begin{center}
\begin{tabular}{|c|c|c|}\hline
PRTA & Main Group & N$_{\text{lines}}$ \\
\hline
1    & [Mg]3p$^2$ 3d$^5$ & 9.9x10$^5$ \\
2    & [Mg]3p$^1$ 3d$^5$ & 1.7x10$^5$ \\
3    & [Mg]3p$^0$ 3d$^5$ & 5470 \\
\hline
\end{tabular}
\caption{Main group and number of lines of the PRTA calculations of example 2.}
\label{table_ex2}
\end{center}
\end{table}  

\subsubsection{Example 3}

In this last example, the initial configuration is [Mg]3p$^4$ 3d$^4$ and the transition is 3s-4p. The DLA calculation contains 9.4x10$^5$ lines. The passive subshell with the smallest contribution to the variance is 3d. In PRTA 1 and 2, one and four electron(s) are removed from the 3d subshell respectively. Table \ref{table_ex3} shows the main group and the number of lines of the PRTA calculations. This transition array reveals two main features. If all the electrons in the 3d subshell are removed (PRTA 2), the width of the main structure (around 0.317 keV) is too large and the structure at higher energy is a little bit shifted. If only one electron is removed, the two structures of the transition array and the width of the main one are in good agreement with the exact DLA calculations.

\begin{table}
\begin{center}
\begin{tabular}{|c|c|c|}\hline
PRTA & Main Group & N$_{\text{lines}}$ \\
\hline
1    & [Mg]3p$^4$ 3d$^3$ & 3.5x10$^5$ \\
2    & [Mg]3p$^4$ 3d$^0$ & 129 \\
\hline
\end{tabular}
\caption{Main group and number of lines of the PRTA calculations of example 3.}
\label{table_ex3}
\end{center}
\end{table}  

\begin{figure}[h]\begin{center}
\includegraphics[scale=0.35]{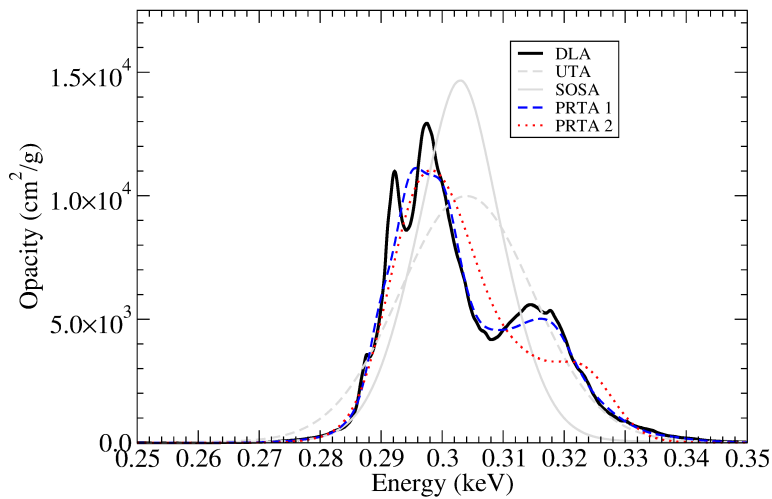}
\caption{(Color online) Comparison between the DLA and the PRTA calculations of the 3s-4p transition with an initial configuration of [Mg]3p$^4$ 3d$^4$ (example 3). The UTA and SOSA statistical modelings are also displayed.}
\label{PRTA_ex3}
\end{center}\end{figure}

\section{Some comparison with experiments}

In this section we compare some calculated spectra with experimental ones. The first and second parts concern absorption and emission experiments, respectively.

\subsection{Photoabsorption experiments}

\begin{figure}[h]\begin{center}
\includegraphics[scale=0.35]{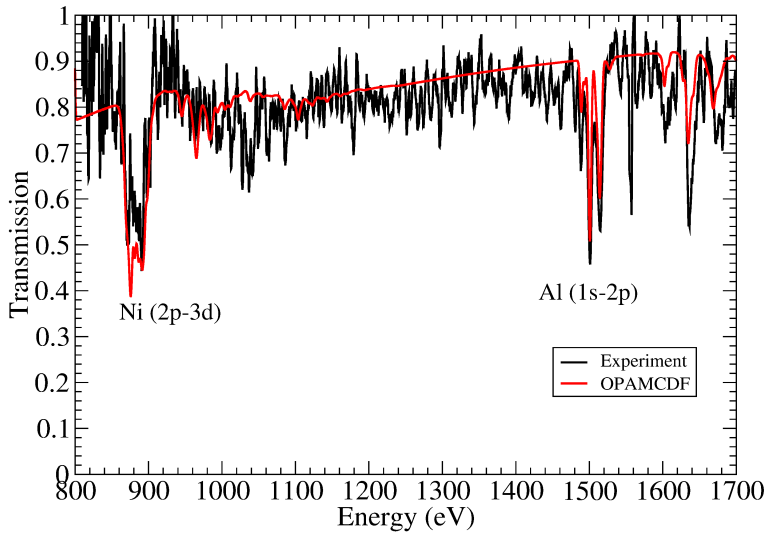}
\caption{(Color online) Comparison of the experimental and calculated transmissions of nickel and aluminum measured at the LULI 2000 facility \cite{Dozieres15}.}
\label{LULI2000_Shot64}
\end{center}\end{figure}

The first comparison is made on an absorption type experiment performed on the LULI 2000 facility. Informations about this experiment can be found in Ref. \cite{Dozieres15}. Two gold cavities are heated by two nanosecond laser beams of 300 J and 1.5 ns pulse duration at 2$\omega$. A thin foil is inserted at 45$^\circ$ between them and is probed by a backlighter produced by a third laser beam impinging on a gold target. Figure \ref{LULI2000_Shot64} presents the comparison of the OPAMCDF results with the experimental transmission spectrum of the shot 64. The agreement in the overall spectral region is found to be good in particular for the 2p-3d structures of nickel and the 1s-2p lines of aluminum. The experimental spectrum being affected by the background noise, the 2p-4d transition of Ni is not clearly visible. In ref. \cite{Dozieres15}, this spectrum was also interpreted with the SCO-RCG code \cite{Pain15}, and the differences between the two calculations are very small.

\begin{figure}[h]\begin{center}
\includegraphics[scale=0.35]{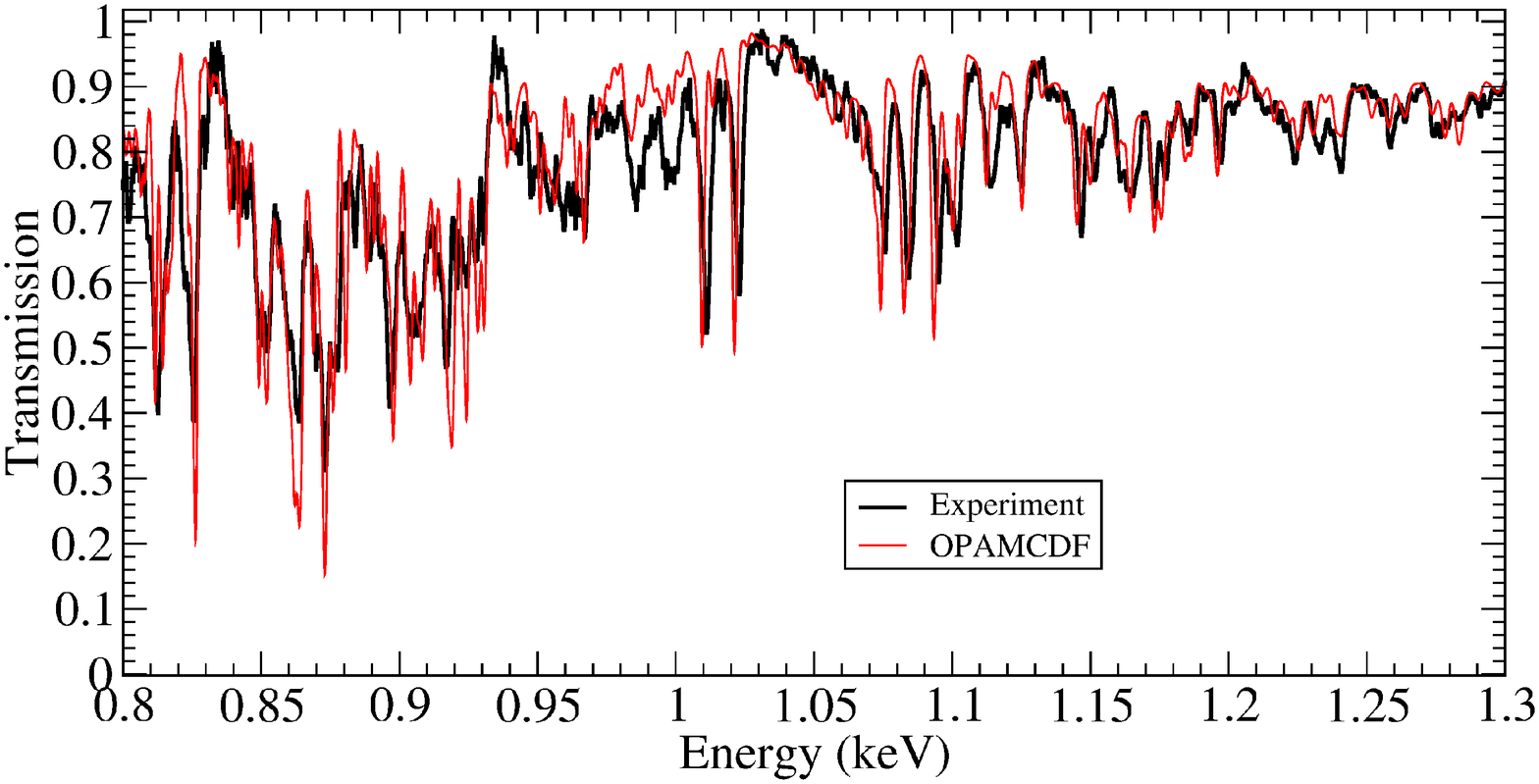}
\caption{(Color online) Comparison of the experimental and calculated transmission spectra of Fe at 150 eV and 0.058 g/cm$^{3}$ measured on the Z facility \cite{Bailey07}.}
\label{Bailey}
\end{center}\end{figure}

The code is also compared to the iron transmission measured on Z-pinch in 2007 using the dynamic hohlraum X-ray source \cite{Bailey07}. An electron temperature and density of 156 $\pm$ 6 eV and 6.9 $\pm$ 1.7$\times 10^{21}$ cm$^{-3}$ were reached. Figure \ref{Bailey} displays the comparison of the OPAMCDF results calculated at 150 eV and 0.058 g/cm$^{3}$ with the experiment using the thin sample of 3.2$\times 10^{-5}$ g.cm$^{-2}$ areal mass. In the computation, the bound-bound opacity is constituted of only DLA computations ; no statistical contribution (PRTA and SOSA) is present in the spectrum. An instrumental resolution power $E/\Delta E$ of 700 is taken into account. The agreement is quite good with the experimental spectrum in the whole energy range.

\subsection{Emission experiment}

\begin{figure}[h]
\begin{center}
\includegraphics[scale=0.35]{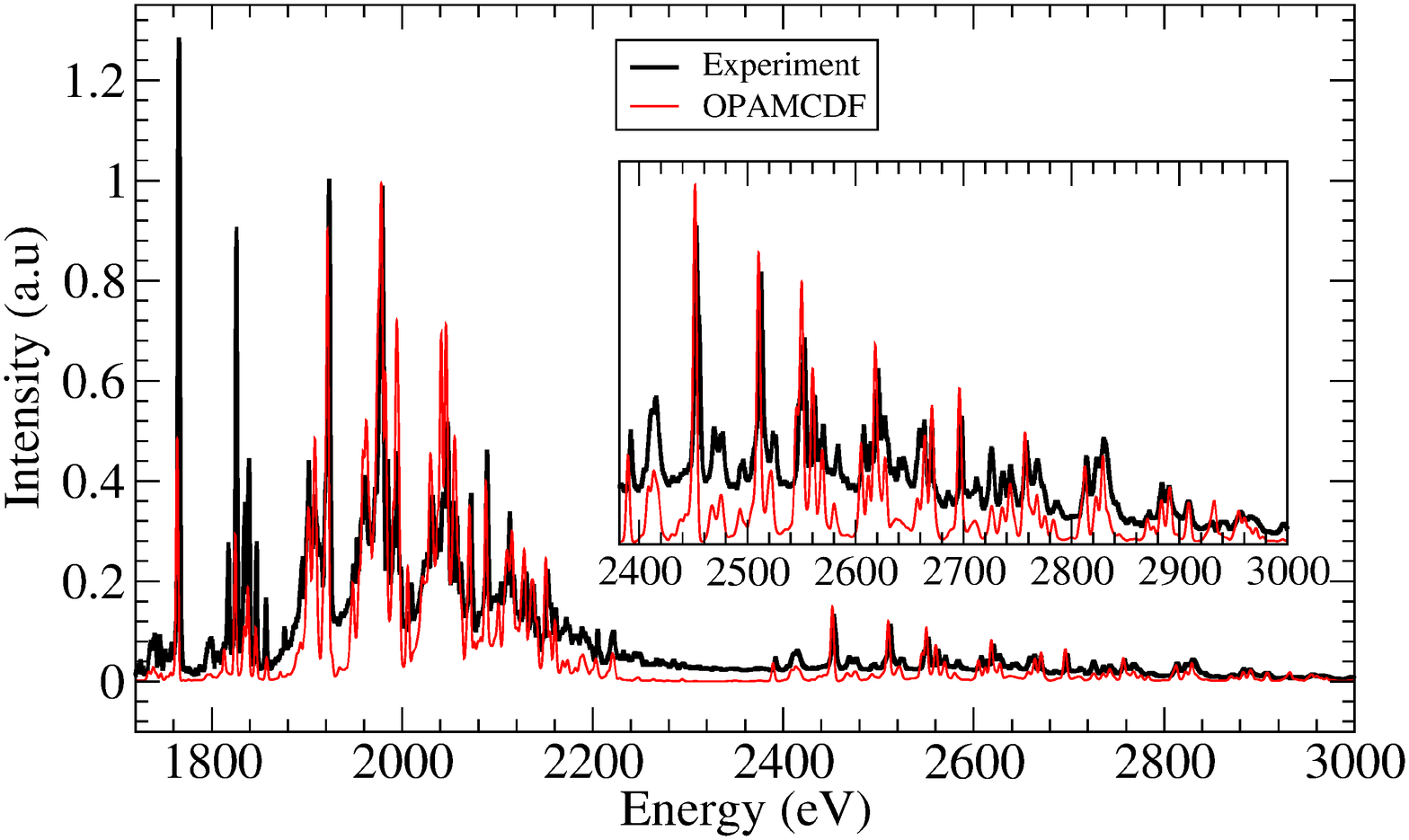}
\caption{(Color online) Comparison of the experimental and calculated emission spectra of Rb measured on the PHELIX laser \cite{Denis-Petit14}.}
\label{Phelix_Rb}
\end{center}
\end{figure}

\begin{figure}[h]\begin{center}
\includegraphics[scale=0.35]{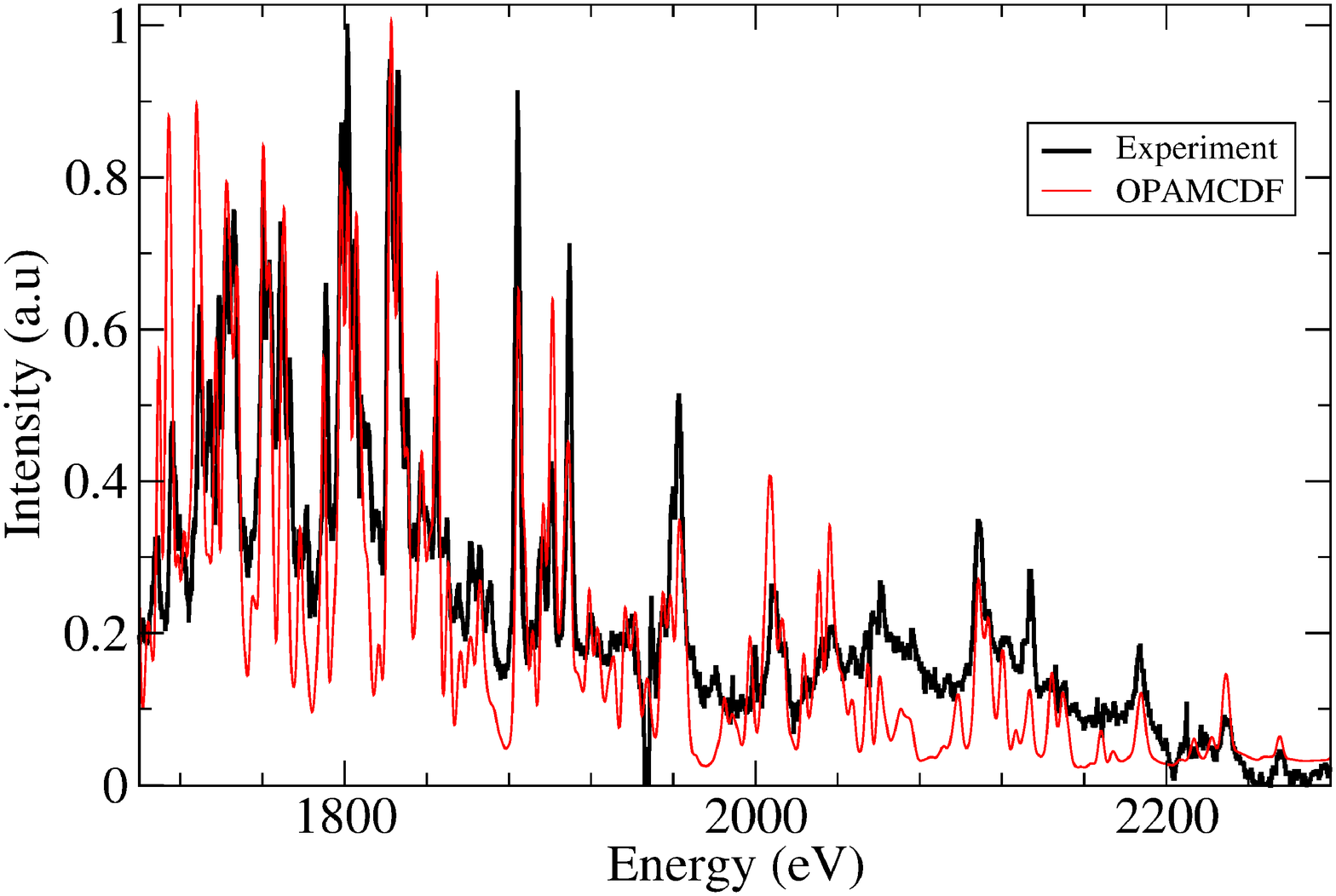}
\caption{(Color online) Comparison of the experimental and calculated emission spectra of Cu measured on the PHELIX laser \cite{Comet17}.}
\label{Phelix_Cu}
\end{center}\end{figure}

Experimental emission spectra of rubidium and copper were measured at the PHELIX facility. Every detail of the experimental setup can be found in \cite{Denis-Petit14}. The plasma was created by a $2\omega$ laser beam of 150 J and 1.4 ns pulse duration. The intensity on the target was estimated to be around 6x10$^{14}$ W/cm$^2$. Comparison between results of the OPAMCDF calculation and experimental spectra are displayed in figure \ref{Phelix_Rb} and \ref{Phelix_Cu} for copper and rubidium targets respectively. The LTE temperature was determined by hydrodynamic simulations and was found to be 270 eV at a density of 10$^{-2}$ g/cm$^3$. The analysis procedure and identification of all the lines can be found in Ref. \cite{Denis-Petit14,Comet17}.

As seen on figures \ref{Phelix_Rb} and \ref{Phelix_Cu}, the calculated spectra are in good agreement with the experimental ones. As described in Ref. \cite{Comet17}, the calculation is made with only one equivalent LTE temperature. Therefore, the few disagreements that can be seen on line ratios for example are due to non-LTE effects plus temperature and density gradients. 

\section{Cross-sections and rates calculations with MCDF}

For spectroscopy of non-LTE plasma, cross-sections and rates are needed for all atomic processes that can populate or depopulate atomic states, levels, configurations or superconfigurations. Additionally to photoexcitation, the MCDF code can be used to calculate cross-sections and rates for photoionization, autoionization, collisional-excitation and ionization using the Distorted-Wave approximation (DW) or the Close Coupling method (except for collisional-ionization). Originally the code was able to calculate rates and cross-sections only between levels. Recent modifications have been added in order to implement the configuration average approximation for all processes listed above.

For collisional processes, the Breit (B), Generalized Breit (GB) and since more recently the Generalized Breit plus the Imaginary part of the matrix element (GBI) can be included in the collisional matrix element. Their expressions can be found in \cite{Sampson09}. For autoionization, the Breit interaction can also be included.

Table \ref{Fe_CS} (\ref{U_CS}) presents the collisional strength of Fe$^{24+}$ (U$^{90+}$) for a scattered electron energy of 70 eV (1 keV). In table \ref{U_ICS}, collisional-ionization cross-sections of U$^{90+}$ and U$^{91+}$ are presented for an electron impact energy of 198 keV. All values are compared with the ATOMIC code results \cite{Sampson09}. This comparison shows that our results differs from the ATOMIC values by less than 10$\%$.

\begin{table}
\begin{center}
\begin{tabular}{|c|c|c|}
\hline
Fe$^{24+}$ & ATOMIC \cite{Sampson09} & MCDF \\
\hline
(1s$^2$-1s2s)$_{0}$ & 7.687 & 8.284 \\
(1s$^2$-1s2s)$_{1}$ & 3.626 & 3.562 \\
(1s$^2$-1s2p$_{1/2}$)$_{0}$ & 2.267 & 2.319 \\
(1s$^2$-1s2p$_{1/2}$)$_{1}$ & 8.079 & 8.217 \\
(1s$^2$-1s2p$_{3/2}$)$_{1}$ & 21.22 & 21.41 \\
(1s$^2$-1s2p$_{1/2}$)$_{2}$ & 10.65 & 10.90 \\
\hline
\end{tabular}
\caption{Collision strengths on Fe$^{24+}$ (in units of 10$^{-4}$). Results are compared with the ATOMIC code \cite{Sampson09}.}
\label{Fe_CS}
\end{center}
\end{table}  

\begin{table}
\begin{center}
\begin{tabular}{|c|c|c|c|c|}
\hline
\multirow{2}{*}{U$^{90+}$} & \multicolumn{2}{c|}{C} & \multicolumn{2}{c|}{C+B} \\
\cline{2-5}
          &    ATOMIC     &        MCDF       &       ATOMIC        &   MCDF   \\
\hline
(1s$^2$-1s2s)$_{0}$ & 15.03 & 15.43 & 23.21 & 23.78 \\
(1s$^2$-1s2s)$_{1}$ & 5.531 & 5.451 & 8.456 & 8.313 \\
(1s$^2$-1s2p$_{1/2}$)$_{0}$ & 3.383 & 3.397 & 1.198 & 1.212 \\
(1s$^2$-1s2p$_{1/2}$)$_{1}$ & 11.94 & 12.03 & 18.15 & 18.01 \\
(1s$^2$-1s2p$_{3/2}$)$_{1}$ & 11.56 & 11.66 & 9.763 & 9.489 \\
(1s$^2$-1s2p$_{1/2}$)$_{2}$ & 6.474 & 6.549 & 9.128 & 9.203 \\
\hline
\end{tabular}
\caption{Collision strengths (in units of 10$^{-5}$) on U$^{90+}$ including only the Coulomb interaction (C) and the Coulomb plus Breit interaction (C+B). Results are compared with those of the ATOMIC code \cite{Sampson09}.}
\label{U_CS}
\end{center}
\end{table}  

\begin{table}
\begin{center}
\begin{tabular}{|c|c|c|c|c|c|}
\hline
\multirow{2}{*}{} 	& \multicolumn{2}{c|}{C} & \multicolumn{2}{c|}{C+GBI} & \multirow{2}{*}{Exp. \cite{Marrs94}}\\
\cline{2-5}
          &    a  &   b  &    a  &   b &  \\
\hline
U$^{90+}$ & 1.97 & 1.95 & 2.96 & 2.94 & 2.82 $\pm$ 0.35\\
U$^{91+}$ & 0.94 & 0.93 & 1.41 & 1.40 &  1.55 $\pm$ 0.27\\
\hline
\end{tabular}
\caption{Collisional-ionization cross sections (in units of 10$^{-24}$ cm$^2$) on U$^{90+}$ and U$^{91+}$ using only the Coulomb interaction (C) and the Coulomb plus the GBI interaction (C+GBI) calculated with MCDF (a) and compared with the ATOMIC code (b) \cite{Sampson09}. The incident electron energy is 198 keV.}
\label{U_ICS}
\end{center}
\end{table}  

\begin{figure}[h]
\begin{center}
\includegraphics[scale=0.35]{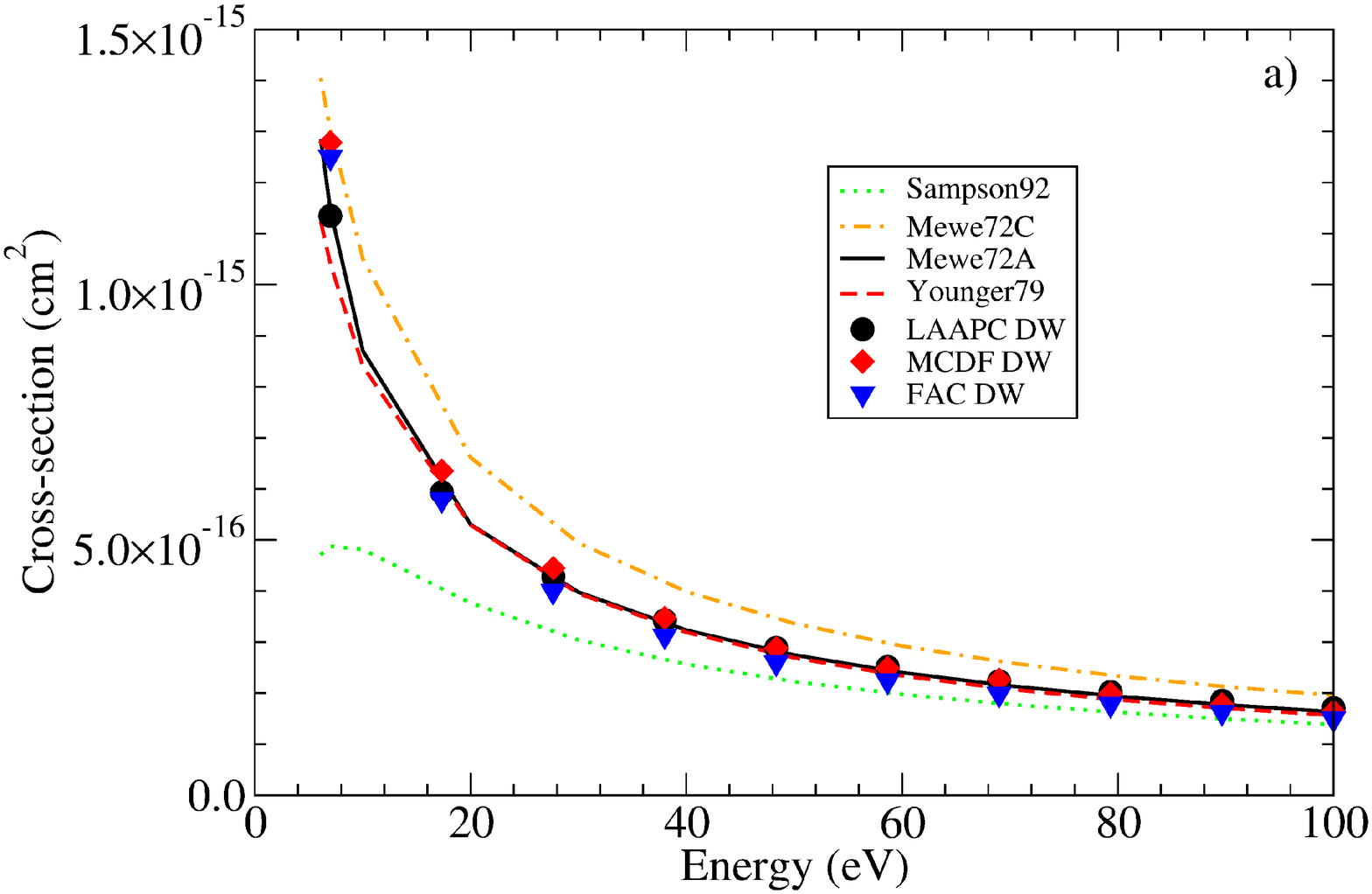}
\includegraphics[scale=0.35]{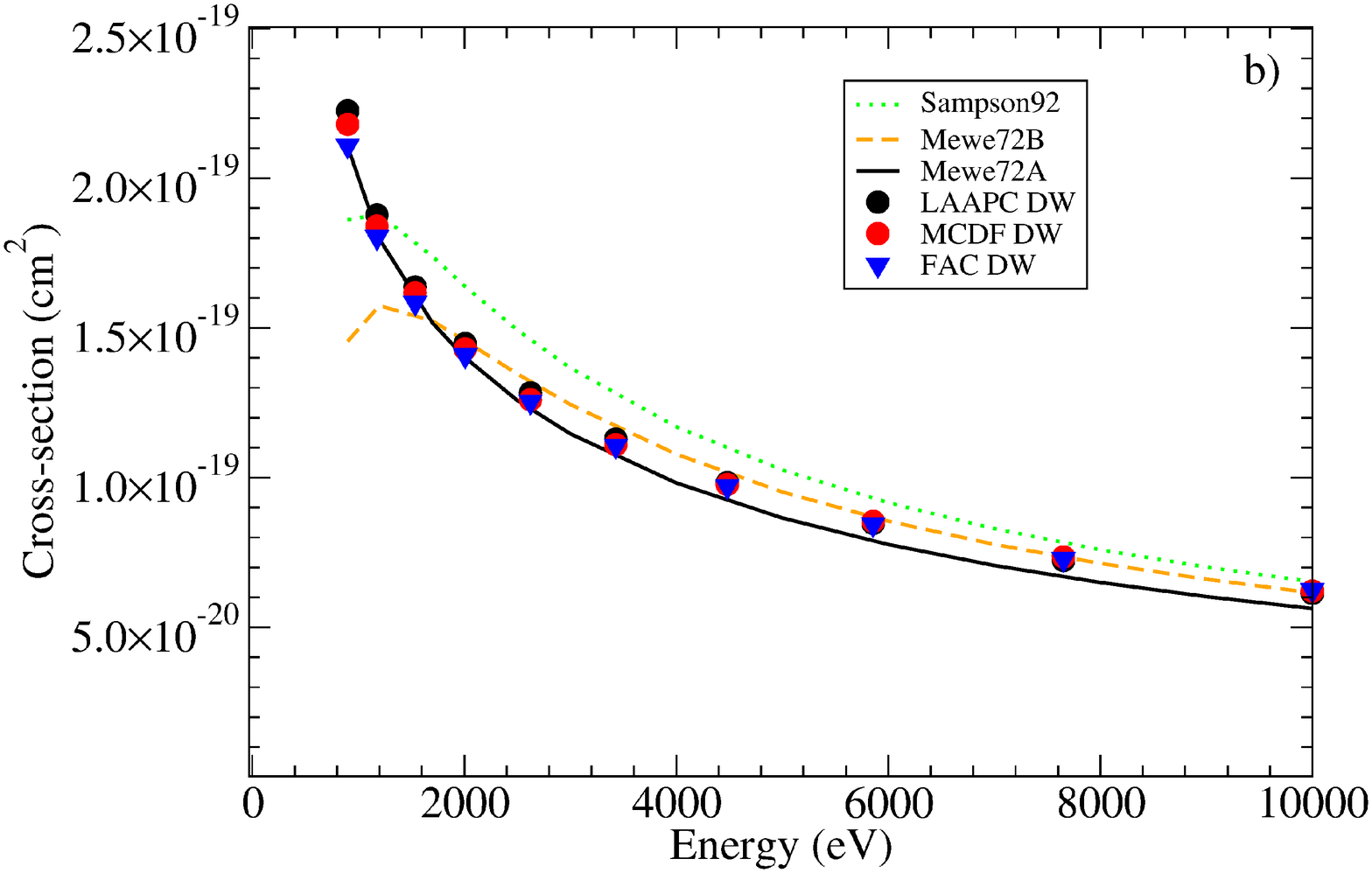}
\caption{(Color online) Comparison of Distorted-Wave collisional-excitation cross-sections in the configuration-average approximation (MCDF, the Los Alamos Atomic Physics code \cite{LosAlamosPhysicsCode}, called LAAPC in the figure , and FAC \cite{Gu08}) with Van Regemorter formula for B$^{2+}$ 2s-2p (a) and Fe$^{16+}$ 2p-3d (b). See text for the Gaunt factors expressions.}
\label{CS_B_Fe}
\end{center}
\end{figure}

The implementation of configuration-averaged cross-sections and rates gives us the opportunity to check the validity of semiempirical formula that are widely used for collisional-radiative modeling. As example, in Figure \ref{CS_B_Fe} a) and b), we show comparisons between Distorted-Wave computations of collisional-excitation cross-sections and Van Regemorter cross-sections \cite{VanRegemorter62}, including different Gaunt factors, for B$^{2+}$ 2s-2p (where a study can be found in \cite{Griem97}), and Fe$^{16+}$ 2p-3d respectively. Our results are labeled MCDF DW and are compared with the Los Alamos Atomic Physics Codes (ACE code) \cite{LosAlamosPhysicsCode} and the Flexible Atomic Code (FAC) \cite{Gu08}. The Gaunt factors $\bar{g}(\varepsilon )$ where $\varepsilon = E/\Delta E$ is the ratio of the electron incident energy $E$ over the threshold energy of the transition $\Delta E$, are based on:

\begin{enumerate}
\item ref. \cite{Sampson92} where the authors have corrected the Gaunt factor of ref. \cite{VanRegemorter62} for high electron energies (labeled Sampson92):

\begin{equation}
\bar{g}(\varepsilon) = 0.2 + \frac{\sqrt{3}}{2\pi} \log \varepsilon
\end{equation}

\item ref. \cite{Mewe72} in which the authors used a Gaunt factor of the form (labeled Mewe72A):

\begin{equation}
\bar{g}(\varepsilon) = A + B\varepsilon^{-1} + C\varepsilon^{-2} + D\log \varepsilon
\end{equation}

\noindent where A, B, C and D depend on the transition involved and the electronic sequence. The values are for Li-like 2s-2p transitions: 

$$A=0.7(1-0.5x), B=1-0.8x, C=-0.5(1-x), D=0.28$$

\noindent with $x=(Z-3)^{-1}$. In the B$^{2+}$ case, $x=0.5$. For Fe$^{16+}$ 2p-3d excitation:

$$A=0.05, B=0.2, C=0, D=0.28$$
 
A generalization is proposed for allowed (electric-dipole) transitions for $\Delta n \neq 0$ (labeled Mewe72B):

\begin{equation}
\bar{g}(\varepsilon) = 0.15 + \frac{\sqrt{3}}{2\pi}\log \varepsilon
\end{equation}

\noindent and for $\Delta n = 0$ (labeled Mewe72C):

\begin{equation}
\bar{g}(\varepsilon) = 0.6 + \frac{\sqrt{3}}{2\pi}\log \varepsilon
\end{equation}

\item ref. \cite{Younger79} where the authors give a Gaunt factor for $\Delta n=0$ transitions (labeled Younger79):

\begin{equation}
\bar{g}(\varepsilon) = \left( 1 - \frac{1}{Z} \right) \left( 0.7 + \frac{1}{n_i} \right) \left(0.6 + \frac{\sqrt{3}}{2\pi} \log{\varepsilon} \right)
\end{equation}

\noindent with $Z$ the spectroscopic atomic charge and $n_i$ the principal quantum number of the initial shell of the transition.

\end{enumerate}

All the Distorted-Wave calculations are very close to each other, especially at high energy when the free wave functions are converging to plane waves. When the electron impact energy decreases, the differences increase. Indeed, in this region, the free wavefunctions are obtained by solving the Dirac equation in a potential $V(r)$. We plan to make Close Coupling calculations in order to check the validity of the Distorted-Wave approximations close to the threshold energy. Concerning the semiempirical Van Regemorter cross sections with different Gaunt factors, we see that, for the 2s-2p excitation on B$^{2+}$, the Mewe72A and the Younger79 Gaunt factor agree very well with the Distorted-Wave calculations. However, the Mewe72C differs from the latter by 20$\%$. The Sampson92 Gaunt factor is not suitable for $\Delta n=0$ transition. For the Fe$^{16+}$ 2p-3d excitation, again the Mewe72A works very well and the agreement is quite good with Mewe72B, for energies greater than 2000 eV. The Sampson92 Gaunt factor differs from the Distorted-Wave computations by $10-20\%$.

\clearpage

\section{Conclusion}

In this paper, we have presented a project dedicated to plasma spectroscopy using the MCDF code initially developed by J. Bruneau. For plasma at local thermodynamic equilibrium, we developed a code, called OPAMCDF, to compute opacity or emissivity spectra. The configuration selection is based on the results from a Relativistic Average Atom Model. The bound-bound opacity consists of DLA, PRTA and SOSA calculations depending on the number of lines in the transition array. A refinement of the PRTA method is proposed by allowing to remove one electron by one electron in the passive subshells instead of removing all of them. Transition arrays computed with this improved version yield a better agreement with the DLA calculations. However, generally the maximum number of lines in the DLA calculations is sufficiently high ($\geq 2\times 10^6$) to ensure that the statistical part of the bound-bound component is negligible. Some comparisons with absorption and emission experimental spectra were discussed.

In the last part of this paper we have presented the capability of this MCDF code to compute cross-sections and rates required for Non-LTE modeling. Comparisons with results from the ATOMIC code show a good agreement. The recent inclusion of configuration-average cross-sections and rates will help us to test some semiempirical cross-sections and rates that are widely used in Non-LTE computations. Examples were presented on the B$^{2+}$ 2s-2p and Fe$^{16+}$ 2p-3d collisional-excitation. 

The perspectives of the project can be divided in two parts. For LTE plasmas, we aim to calculate detailed photoionization of low-Z elements. Moreover, the code is able to work whatever the transition multipolarity. A study is needed in order to see the impact of electric and magnetic multipole lines and photoionization edges on an opacity spectrum. We plan to add some collisional and autoionization widths using detailed or configuration-averaged cross-sections. Concerning the cross-sections computations for Non-LTE plasmas spectroscopy, we plan to perform some Close Coupling calculations to check the validity of the Distorted-Wave approximation for free electrons with low impact energy. Furthermore, the top-up contributions to the collisional-excitation need to be included using the Kummer transformation or the Coulomb-Bethe approximation \cite{Sampson09} in the MCDF code to take into account excitation channels in which the free electron has high angular momenta.

\section{Acknowledgments}

We are indebted to J. Bruneau for the transmission of the MCDF code. We wish to thank also D. Denis-Petit et al. \cite{Denis-Petit14} for the experimental spectrum measured on the Phelix laser, J. Bailey et al. \cite{Bailey07} for the spectrum measured on the Z-pinch at the Sandia National Laboratory and M. Dozi\`eres et al. \cite{Dozieres15} for the experimental spectrum measured on the LULI 2000 facility.

\end{document}